\documentclass{article}
\usepackage{amsmath}
\usepackage{amssymb}
\usepackage{setspace}
\begin{document}
\title{Subtlety in the Use of Maxwell's Equation and a New Electromagnetic
Wave in Electron Plasmas}
\author{H. Saleem\\National Centre for Physics, Quaid-i-Azam University Campus, \\Islamabad, Pakistan}
\date{}
\maketitle
\begin{abstract}

The ambiguity involved in the use of Maxwell's equation particularly
in electron plasmas is discussed. It is pointed out that in the slow
time scale perturbations the displacement current is ignored but it
does not imply that the electron density fluctuations vanish. The
contradictions in the assumptions and approximations used in the
literature on this subject are discussed. A new low frequency
electromagnetic wave is described which is a normal mode of
non-uniform magnetized electron plasmas. This wave can couple with
plasma hybrid oscillations if ion dynamics is taken into account. It
is stressed that the electron magnetohydrodynamics (EMHD) model
seems to be simple but in fact its use is subtle and its scope is
very limited.
\end{abstract}

\large{

It is well-known that the Maxwell's equation reduces to Ampere's law
when displacement current is neglected in the slow time scale
phenomena. The divergence of current vanishes and hence several
researchers conclude that the density fluctuations in a charged
particle system disappears. In electron-ion plasma the
quasi-neutrality is used as for example in the case of Alfven waves
but it is justified. On the other hand, if ions are considered to be
stationary, then the electron density fluctuations do not appear due
to Ampere's law in previous investigations both in unmagnetized [1 -
6] and magnetized [7 -9] plasmas and here it is incorrect. The forms
of transverse waves in cartesian geometry have serious flaws.
\\A great deal of literature exists on the topic of magnetic field
generation on laser-plasma [1 -6] and cosmological [10] scales
assuming the system to be unmagnetized initially. On the other hand,
the efforts have also been made to find out some mechanism for the
generation of magnetic fluctuations parallel to external magnetic
field which can be very important in plasma opening switches (PoS)
[7, 8]. Thermomagnetic instability in laser plasmas has also been
investigated assuming electron density to be fixed in magnetized
inhomogeneous plasma [9]. \\It is a misconception that the zero
displacement current means the zero density fluctuation in electron
plasmas. Based on this conclusion drawn from Ampere's law several
research papers and review articles have appeared in journals and
books which need to be corrected.\\For the study of magnetic field
generation and plasma switches a simpler single fluid model called
electron magnetohydrodynamics (EMHD) was presented [11]. In this
model the electron inertia term in equation of motion is neglected
and Ampere's law is used assuming the time scale
$|\partial_{t}|<<\omega_{pe},ck$. Moreover, the ions are assumed to
be static in the limit $\omega_{pi}<<|\partial_{t}|$ where
$\omega_{pj}=(4\pi n_{0}e^{2}/m_{j})^{\frac{1}{2}}$ is the plasma
oscillation frequency of jth species, $k$ is wave number and $c$ is
velocity of light. In magnetized plasmas, EMHD is supposed to be
valid for $\Omega_{i}<<|\partial_{t}|<<\Omega_{e}$ where
$\Omega_{j}=eB_{0}/m_{j}c$ is the gyrofrequency of the jth species.
\\The EMHD model contains several discrepancies and contradictions.
But it has been continuously used for the last several decades. The
case of magnetized plasmas is more important because of its
laboratory applications. The previous concept of negligible density
fluctuations associated with low frequency perturbation in the
electron plasma is still being followed [12]. In some works the
nonlinear whistler wave in pure electron plasmas have been studied.
Since these waves are basically pure transverse in nature,
therefore, the electron density perturbations in nonlinear stage may
be neglected [13], but this point also needs to be checked
carefully.
\\Let us choose the wave geometry the same as was considered in the
so called magnetic drift wave (MDW) in Refs. [7, 8]. The constant
external magnetic field is
$\mathbf{\hat{B}}_{0}=B_{0}\mathbf{\hat{z}}$, the density gradient
is $\nabla n_{j0}=-\mathbf{\hat{x}}\frac{dn_{0}}{dx}$ and
perturbation is proportional to $e^{i(k_{y}y-\omega t)}$ while the
subscript naught (0) denotes equilibrium quantities. The only
difference is that we assume the density gradient along negative
x-axis in resemblance with the well-known drift wave case while in
the derivation of the dispersion relation for MDW it was assumed
along positive x-axis.
\\Now we analyse the physical model in detail based on EMHD equations which contains fundamental errors.
In the limit $\Omega_{i}<<\omega$ the ions are assumed to be
stationary and for $\omega<<\omega_{pe},{c}{k}, \Omega_{e}$, the
displacement current is neglected in the Maxwell's equation,
$$\nabla\times\textbf{B}=e\frac{4\pi}{c}\textbf{J}+\frac{1}{c}\partial_{t}\textbf{E}\eqno(1)$$
which then becomes Ampere's law,
$$\nabla\times\textbf{B}=\frac{4\pi}{c}\textbf{J}\eqno(2)$$ In the case of pure
electron plasma we have $\textbf{J}_{1}=-en_{e0}\textbf{v}_{e1}$
where the subscript one (1) denotes linearly perturbed quantities.
Since Ampere's law implies $\nabla.\textbf{J}_{1}=0$, therefore in
the derivation of dispersion relation of MDW it is deduced that
$n_{e1}=0$. Using the limit $|\partial_{t}|<<\Omega_{e}$, electron
acceleration term is also neglected in the equation of motion which
then becomes very simple as,
$$\textbf{E}_{1}=\frac{c}{B_{0}}\textbf{v}_{e1}\times\mathbf{\hat{z}}\eqno(3)$$
The magnetic fluctuation is assumed to be along z-axis and hence the
Faraday's law
$$\nabla\times\textbf{E}_{1}=-\frac{1}{c}\partial_{t}\textbf{B}_{1}\eqno(4)$$
requires $\textbf{E}_{1}=E_{1}\mathbf{\hat{x}}$ which implies
$\nabla.\textbf{E}_{1}=0$ and therefore the wave seems to be pure
transverse. Equation (2) gives,
$$\nabla.\textbf{v}_{e1}=\frac{c}{4\pi en_{0}}\frac{\nabla
n_{0}}{n_{0}}.(\nabla\times\textbf{B}_{1})\eqno(5)$$ and hence
$\nabla.\textbf{v}_{e1}\neq0$. \\The final dispersion relation under
local approximation turns out to be [8],
$$\omega=\lambda^{2}_{e}k^{2}_{y}\left(\frac{\kappa_{n}}{k_{y}}\Omega_{e}\right)=k_{y}v_{A}\left(\frac{c}{\omega_{pi}L_{n}}\right)\eqno(6)$$
where $v_{A}=B_{0}/\sqrt{4\pi n_{0} m_e }$ is the electron Alfven
speed, $\omega_{pi}=\left(4\pi
n_{0}e^{2}/m_{i}\right)^{\frac{1}{2}}$ is the ion plasma
oscillation, $L_{n}=1/\kappa_{n}$,
$\kappa_{n}=\left|\frac{1}{n_{0}}\frac{dn_{0}}{dx}\right|$ and
$\lambda_{e}=c/\omega_{pe}$. \\ A very trivial but crucial point has
been overlooked in the above assumptions and approximations. The
similar treatment in case of unmagnetized plasmas has also been
followed by Jones [4] and several others. He also derived a so
called new low frequency transverse electromagnetic wave in
unmagnetized electron plasmas using $\nabla.\textbf{E}_{1}=0$ and
$\nabla.\textbf{v}_{e1}\neq0$. Later on, several linear and
nonlinear investigation of this mode within the local approximation
were performed.
\\Most of these authors have used
the electron magnetohydrodynamics (EMHD) set of equations [11]. The
EMHD model was presented to reduce the time and spacial scales of
two component electron-ion plasma to a single fluid which could
explain some important phenomena. The assumptions and
approximations, which seem to be very reasonable apparently, are in
fact self-contradictory and erroneous. Let us analyse the above
simple linear model and look into the physical picture in some
detail. It will help us in finding out the interesting and important
clear results which will be very useful for future studies and
applications. Equation (5) implies $\nabla.\textbf{v}_{e1}\neq0$ and
Eq. (2) requires, $$\nabla
n_{0}.\textbf{v}_{e1}+\nabla.\textbf{v}_{e1}=0\eqno(7)$$ Thus the
electron velocity has two non-zero components $v_{ex1}$ and
$v_{ey1}$. If it is so, then Eq. (3) yields $E_{1y}\neq0$ and hence
$\nabla.\textbf{E}_{1}\neq0$ which is a contradiction to the initial
assumption that the MDW is pure transverse. Two important points
need attention \begin{enumerate}
\item Ampere's Law does not necessarily imply that electron density fluctuations are zero.
\item Since $\kappa_{n}<<k_{y}$ under local approximation, therefore
$\omega$ can be near lower hybrid oscillations
$(\Omega_{e}\Omega_{i})^{\frac{1}{2}}$. Thus the ion dynamics cannot
be ignored particularly in the case of hydrogen plasma.
\end{enumerate}
First we present a model for pure electron plasma which is more
useful for heavier ion plasmas where the approximation $\omega_{pi},
\Omega_{i}<<\omega$ seems to be more logical and ions can be assumed
to be static. The experiments have been performed to produce pure
pair-ion fullerene plasmas [14, 15, 16] in Japan. However, a
criterion for the pure pair-ion plasma has been defined [17] and it
shows that the fullerene plasma produced in experiments was not pure
pair-ion plasma. On the other hand, it has been learnt that there
are plans to study electron shear flow effects in barium plasma in
USA [18]. Since we are in the lower hybrid range of frequencies
$\Omega_{i}<<\omega<<\Omega_{e}$, therefore we retain the electron
inertia term in equation of motion which for $\textbf{E}=(E_{1x},
E_{1y}, 0)$ yields,
$$v_{e1x}=\frac{e}{m_{e}\Omega^{2}_{e}}\{\iota \omega
E_{1x}+\Omega_{e}E_{1y}\}\eqno(8)$$
$$v_{e1y}=\frac{e}{m_{e}\Omega^{2}_{e}}\{\iota \omega
E_{1y}-\Omega_{e}E_{1x}\}\eqno(9)$$ Then Poisson equation becomes,
$$\left\{\left(1+\frac{\Omega^{2}_{e}}{\omega^{2}_{pe}}\right)\omega+\Omega_{e}\frac{\kappa_{n}}{k_{y}}\right\}\iota
k_{y}=\Omega_{e}E_{1x}\eqno(10)$$ and Ampere's law yields,
$$E_{1y}=(\lambda^{2}_{e}k^{2}_{y})\frac{\Omega_{e}}{\omega}\iota
E_{1x}\eqno(11)$$ Then Eqs. (10) and (11) give a new low (or hybrid)
frequency electromagnetic wave in non-uniform electron plasmas as,
$$\omega=-\frac{\lambda^{2}_{e}k^{2}_{y}}{\left\{1+\lambda^{2}_{e}k^{2}_{y}(1+\frac{\Omega^{2}_{e}}{\omega^{2}_{pe}})
\right\}}\left(\frac{\kappa_{n}}{k_{y}}\Omega_{e}\right)\eqno(12)$$
The term
$\mu=\lambda^{2}_{e}k^{2}_{y}\left(1+\frac{\Omega^{2}_{e}}{\omega^{2}_{pe}}\right)$
is crucial. The factor $\lambda^{2}_{e}k^{2}_{y}$ appears due to
electron inertia term in equation of motion and
$\lambda^{2}_{e}k^{2}_{y}\left(\frac{\Omega^{2}_{e}}{\omega^{2}_{pe}}\right)$
appears due to density fluctuations (or compressibility) in Poisson
equation. In heavier ion plasmas Eq. (12) can be very important and
even in hydrogen plasma it can be valid if
$\kappa_{n}/k_{y}\sim(10^{-1})$,
$\lambda^{2}_{e}k^{2}_{e}\sim(1)$ and
$\Omega^{2}/\omega^{2}_{pe}\lesssim1$ so that $\Omega_{i}<<\omega$
remains valid. However, in hydrogen plasmas, the frequency of this
mode will be closer to $(\Omega_{e}\Omega_{i})^{\frac{1}{2}}$ and
therefore it is preferable to include ion dynamics. Ion momentum
equation gives,
$$v_{i1x}=\frac{e}{m_{i}}\frac{1}{(\omega^{2}-\Omega^{2}_{i})}(\iota
\omega E_{1x}-\frac{e}{m_{i}}E_{1y})\eqno(13)$$
$$v_{i1y}=\frac{e}{m_{i}}\frac{1}{(\omega^{2}-\Omega^{2}_{i})}(\iota
\omega E_{1y}-\Omega_{i}E_{1x})\eqno(14)$$ Then the Poisson equation
becomes,
$$\left[\left(1+\frac{\omega^{2}_{pe}}{\Omega^{2}_{e}}\right)\omega^{3}+\left(\frac{\omega^{2}_{pe}}{\Omega_{e}}\frac{\kappa_{n}}{k_{y}}\right)
\omega^{2}-\omega^{2}_{pi}\omega\right]\iota
k_{y}E_{1y}$$$$=\left[\omega^{2}_{pi}(\Omega_{i}k_{y}-\omega\kappa_{n})+\frac{\omega^{2}_{pe}}{\Omega^{2}_{e}}
(\Omega_{e}k_{y})\omega^{2}\right]E_{1x}\eqno(15)$$ In the
electrostatic case $E_{1x}=0$ it reduces to,
$$\omega^{2}+\frac{\kappa_{n}}{k_{y}}\Omega_{e}-\Omega_{i}\Omega_{e}=0\eqno(16)$$
In the electromagnetic case we obtain,
$$\left\{1+\lambda^{2}_{e}k^{2}_{y}\left(1+\frac{\Omega^{2}_{e}}{\omega^{2}_{pe}}\right)\right\}\omega^{2}+\lambda^{2}_{e}k^{2}_{y}
\left(\frac{\kappa_{n}}{k_{y}}\Omega_{e}\right)\omega-\lambda^{2}_{e}k^{2}_{y}(\Omega_{e}\Omega_{i})=0\eqno(17)$$
The electromagnetic wave described in Eq. (12) couples with lower
hybrid oscillations in Eq. (17) due to ions contribution in the
limit $\omega^{2}\simeq\Omega_{e}\Omega_{i}$,
$\omega\kappa_{n}\simeq\Omega_{i}k_{y}$ and
$\lambda^{2}_{e}k^{2}_{y}\sim(1)$. \\The new electromagnetic wave
described by dispersion relation (12) will have lot of applications
in plasma dynamics for example in plasma switches and in plasma
transport in tokamaks. It is partially longitudinal and partially
transverse. \\To summarize, the ambiguity in the case of Ampere's
law created by a great deal of literature on EMHD models presented
for the generation of magnetic fields has been clarified. It has
been pointed out that when the divergence of transverse current is
zero, it does not mean that longitudinal current is zero. Therefore,
the Ampere's law does not necessarily imply that there are no
density perturbations in the charged particle system. Coupling of
Ampere's law with Poisson equation gives a new partially
longitudinal and partially transverse wave which exists in
non-uniform magnetized electron plasmas. This mode has been
overlooked so far in the plasmas, and in our point of view it plays
important role in plasma dynamics.


\pagebreak

 }

\end{document}